\documentclass[journal]{vgtc}                
\ifpdf
  \pdfoutput=1\relax                   
  \usepackage{graphicx}                
  \DeclareGraphicsExtensions{.pdf,.png,.jpg,.jpeg} 
\else
  \usepackage{graphicx}                
  \DeclareGraphicsExtensions{.eps}     
\fi%
\newcommand{\etal}{{\textit{et~al.}}\xspace}

\usepackage{xspace}

\graphicspath{{figures/}{pictures/}{images/}{./}} 

\usepackage{microtype}                 
\PassOptionsToPackage{warn}{textcomp}  
\usepackage{textcomp}                  
\usepackage{mathptmx}                  
\usepackage{times}                     
\usepackage{cite}                      
\usepackage{tabu}                      
\usepackage{booktabs}                  
\usepackage{amsmath}

\usepackage[ruled,vlined,linesnumbered]{algorithm2e}
\usepackage{enumitem}
\usepackage{xspace}
\usepackage[table,usenames,dvipsnames]{xcolor}
\usepackage{booktabs}
\usepackage{multirow}
\usepackage{array}
\usepackage{cellspace}
\usepackage[10pt]{moresize} 
\usepackage{amsmath}
\usepackage{listings}
\usepackage{xcolor}
\usepackage{amsmath}
\usepackage{MnSymbol}

\onlineid{1090}

\vgtccategory{Research}
\vgtcpapertype{Systems \& Rendering}

\newcommand{\name}{Datamator\xspace}

\title{\name : An Intelligent Authoring Tool for Creating Datamations via Data Query Decomposition}

\author{Yi Guo, Nan Cao, Ligan Cai, Yanqiu Wu, Daniel Weiskopf, Danqing Shi and Qing Chen}
\authorfooter{
\item
Yi Guo, Nan Cao, Ligan Cai, Yanqiu Wu, Danqing Shi and Qing Chen are with Intelligent Big Data Visualization Lab, Tongji University. E-mail: \{2010937, nan.cao, tsailgan, 1941923, sdq, qingchen\}@tongji.edu.cn. \\
Nan Cao \& Qing Chen are the corresponding authors.
\item 
Daniel Weiskopf is with University of Stuttgart. E-mail: \{ Daniel.Weiskopf@visus.uni-stuttgart.de\}
}

\shortauthortitle{Biv \MakeLowercase{\textit{et al.}}: Global Illumination for Fun and Profit}


\CCScatlist{ 
 \CCScat{K.6.1}{Management of Computing and Information Systems}%
{Project and People Management}{Life Cycle};
 \CCScat{K.7.m}{The Computing Profession}{Miscellaneous}{Ethics}
}

\teaser{
  \centering
  \includegraphics[width=\linewidth]{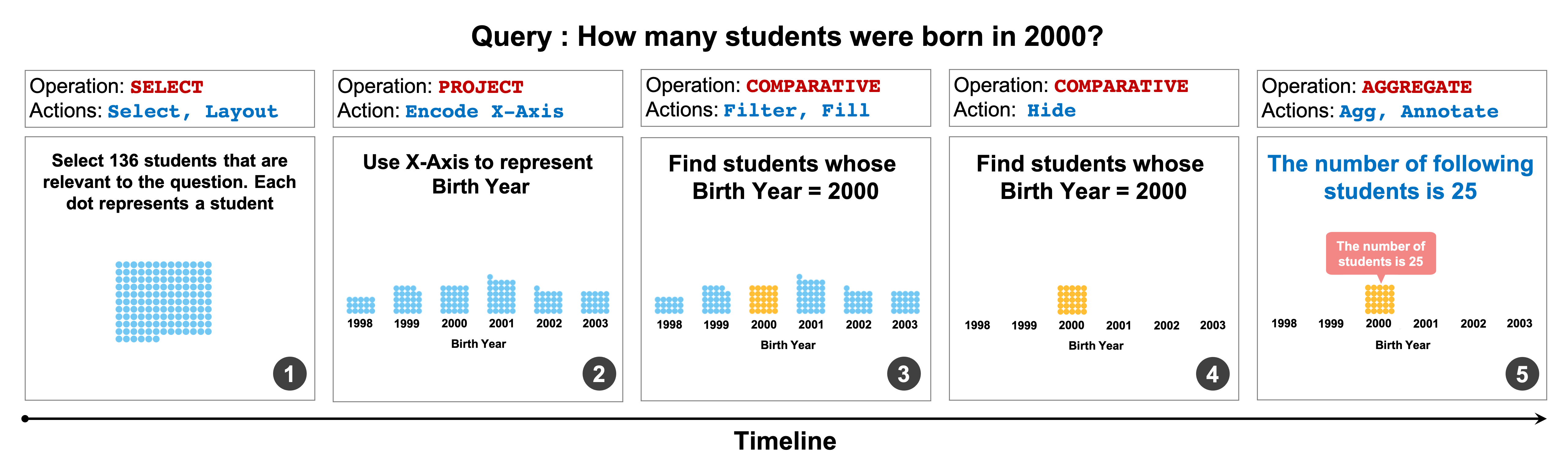}
  \caption{A datamation generated by \name that visualizes the analysis pipeline regarding to the data query ``how many students were born in 2000?" through an action-oriented unit visualization.}
  \label{fig:teaser}
}

\abstract{
Datamation is designed to animate an analysis pipeline step by step, which is an intuitive and effective way to interpret the results from data analysis. However, creating a datamation is not easy. 
A qualified datamation needs to not only provide a correct analysis result but also ensure that the data flow and animation are coherent. Existing animation authoring tools focus on either leveraging algorithms to automatically generate an animation based on user-provided charts or building graphical user interfaces to provide a programming-free authoring environment for users. None of them are able to help users translate an analysis task into a series of data operations to form an analysis pipeline and visualize them as a datamation. To fill this gap, we introduce \name, an intelligent authoring tool developed to support datamation design and generation. It leverages a novel data query decomposition model to allow users to generate an initial datamation by simply inputting a data query in natural language. The initial datamation can be refined via rich interactions and a feedback mechanism is utilized to update the decomposition model based on user knowledge and preferences. Our system produces an animated sequence of visualizations driven by a set of low-level data actions.  It supports unit visualizations, which provide a mapping from each data item to a unique visual mark. We demonstrate the effectiveness of \name via a series of evaluations including case studies, performance validation, and a controlled user study. 

}

\vgtcinsertpkg

\begin{document}



\firstsection{Introduction}
\maketitle
A datamation~\cite{pu2021datamations} is designed to help interpret the results from data analysis by animating the detailed analysis pipeline step by step. Although it has been demonstrated to be an intuitive and effective way of interpreting the analysis, creating a datamation is not easy. The designer needs to decompose a complex data analysis task into a sequence of fundamental data operations and represent the intermediate results via appropriate visualizations that could be smoothly connected through animations. To this end, one needs to acquire multiple skills, including data analysis, visualization, and animation design, which will be a challenge for ordinary users with a little technical background. 

During the past decades, in the field of data visualization, techniques for creating insightful animations have been extensively studied~\cite{ge2020canis,kim2021gemini,shi2021autoclips,bostock2011d3} and a number of authoring tools~\cite{amini2018hooked,ge2021cast,thompson2021data} have also been developed for helping users create smooth transitions between charts. These techniques and tools greatly lower the technical barriers of designing meaningful animated transitions in a data visualization. However, none of them aim to help users translate an analysis task into a series of data operations to form an analysis pipeline and visualize them as a datamation. Designing such a datamation authoring tool is not simple and a number of challenges exist. First, it is usually difficult for a designer with little data analysis knowledge to clearly and precisely describe an analysis task. In most cases, they can only tell what they would like to find from the data in natural language. Second, selecting a series of data operations to create a datamation is hard because the right solution should not only provide the correct analysis result but also needs to ensure a continuous data flow (i.e., the output of the current step is the input of the next step) so that the changes can be smoothly represented in animation. Third, generating a meaningful animation of the analysis pipeline is difficult. Even though the intermediate analysis results may be shown in different forms of visualizations, the animation should be able to smoothly connect them without losing focus or increasing cognitive load.

To address the above issues, in this paper, we introduce \name, the first authoring tool developed for creating datamations. \name employs an advanced deep learning model that can directly and automatically decompose a data query described in natural language into a sequence of pre-defined operations. Each operation corresponds to a series of low-level actions that drive a captioned unit visualization to visualize the change of the intermediate analysis results through animated transitions. A preview of the generated datamation is displayed in an editor in which users can edit the analysis pipeline and the details of each of its analysis actions. The system also incorporates a deep knowledge editing network to help improve the generation results based on users' feedback. In particular, users can calibrate the decomposition results in the datamation editor and feed their modifications back into the decomposition model to improve it. 

The contributions of the paper are as follows:

\begin{itemize}[leftmargin=10pt,topsep=2pt]
\itemsep -.5mm
\item {\bf System.} We introduce the first, to the best of our knowledge, an intelligent authoring tool that is developed to support datamation design and generation. 
With this system, a user can easily generate an initial datamation by simply specifying a data query in natural language and then editing to refine it via rich interactions and a feedback mechanism based on his/her knowledge and preferences.

\item {\bf Data Query Decomposition with Calibration.} We introduce a data query decomposition model that automatically converts a natural language data query into a sequence of data analysis operations for generating a datamation. The model integrates a deep knowledge calibration network to support an efficient feedback mechanism. 

\item{\bf Action-Oriented Unit Visualization.} We introduce a unit visualization driven by a set of low-level data actions designed for illustrating datamations. The visualization uses a caption to illustrate the semantics of each action and show the effects of the action via animated transitions of units.

\end{itemize}

Please note that we adopt the term unit visualization from Drucker and Fernandez~\cite{drucker2015unifying}, which describes the class of visualization techniques that have a direct mapping between individual data items and corresponding unique visual marks. 


We evaluate \name via quantitative experiments of the performance of the query decomposition model, a controlled user study to assess the quality of the generated animations, and interviews with two expert users to verify the usability of the system.
\begin{figure*}[!htb]
\setlength{\abovecaptionskip}{10pt}
\centering 
\includegraphics[width=0.98\textwidth]{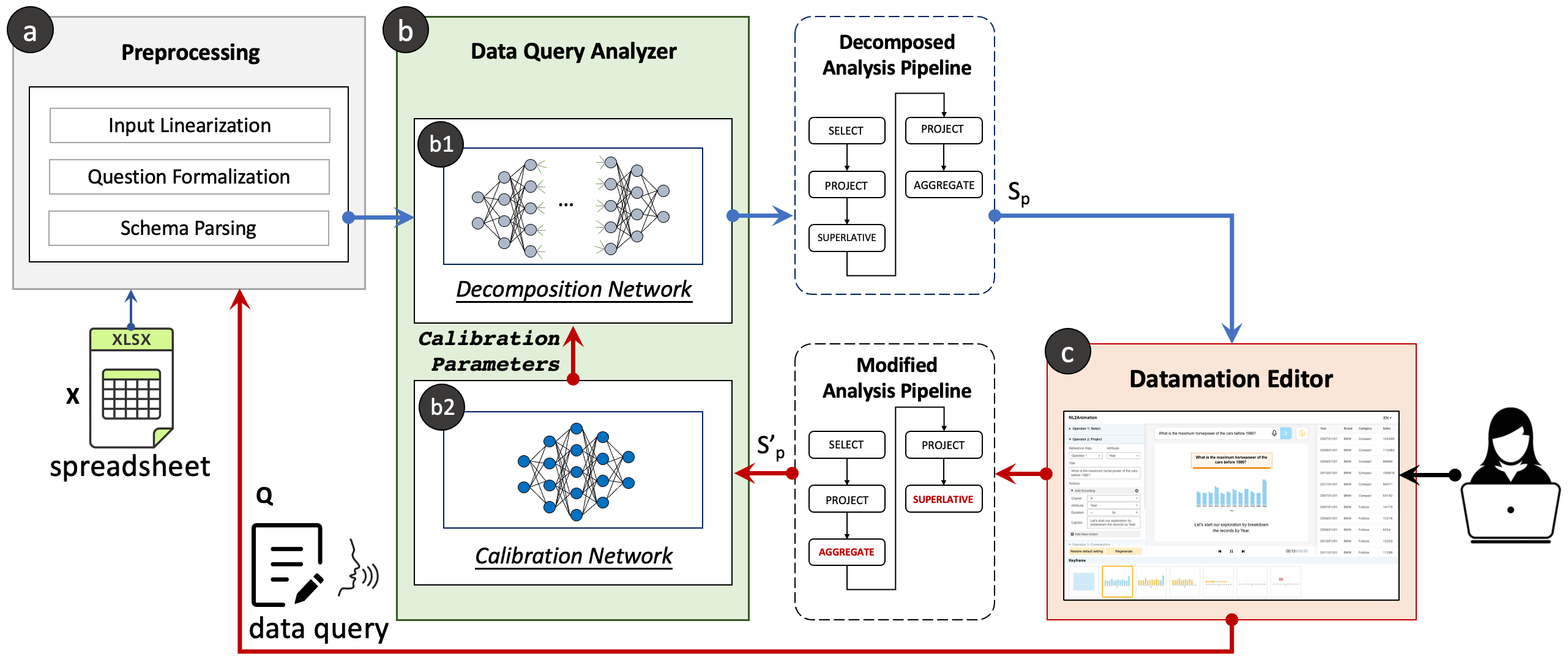}
\caption{The architecture design of \name system consists of three major modules: (a) preprocessing, (b) data query analyzer, (c) datamation editor.
}
\label{fig:pipeline}
\vspace{-1.5em}
\end{figure*}

\section{Related Work}
In this section, we review the recent studies that are most relevant to our work: animation in data visualization, animation generation, and natural language interface for data visualization.

\subsection{Animation in Data Visualization}
In the field of data visualization, animations is usually used for illustrating the change of data~\cite{robertson2008effectiveness}, showing the transitions between visualization views~\cite{heer2007animated}, highlighting relationships~\cite{fisher2010animation,amini2018hooked}, and catching attentions~\cite{robertson2008effectiveness}. It has also been used for supporting data analysis. Some works focus on using animation for highlighting \cite{Ware:2004:MSR} or adding information shown in static plots to boost reader comprehension. 
Hypothetical outcome plots~(HOPs)~\cite{hullman2015hypothetical}, for example, augment static visualizations (e.g., error bars) with animated frames of random draws from the underlying sampling distribution to convey uncertainty.
The most recent work by Pu~\etal~\cite{pu2021datamations} introduces the idea of datamation, which is designed to support users in interpreting the results from complex data analysis by animating the detailed analysis pipeline step by step. 
The datamation enhances a static visualization with details from the data analysis phase, which can convey important insights and help people understand specific analysis results in everyday settings. 

Following the idea of datamation~\cite{pu2021datamations}, we have developed \name, an intelligent authoring tool that supports datamation design and generation. We introduce a unit visualization driven by a set of low-level data actions designed for illustrating datamations. The visualization uses captions to illustrate the semantics of each action and show the effects of the action via animated transitions of units.

\subsection{Animation Generation}
Creating animated visualization can be difficult and time-consuming. A range of tools has been introduced to help users create animated transitions. Comprehensive libraries, such as D3~\cite{bostock2011d3}, allow flexible creation and great expressiveness but require significant effort. Users need to write program code to calculate and assign values for low-level components, impairing the ease of use. High-level grammars for animated transitions can help balance the trade-off between creation flexibility and ease of use. For instance, Gemini and Gemini2~\cite{kim2020gemini,kim2021gemini} perform and recommend animated transitions between two Vega-Lite charts. Canis~\cite{ge2020canis} supports animation effects and temporal functions to selected marks. While these grammars avoid imperative programming, it is still challenging for ordinary users with a little programming background to operate the grammars. 

To provide a programming-free environment, existing approaches choose to either automatically generate animations by algorithms~\cite{li2021anivis,shi2021autoclips} or provide graphical interfaces for authoring~\cite{amini2016authoring,ge2021cast,thompson2021data}. For instances, AutoClips~\cite{shi2021autoclips} takes a sequence of data facts~(e.g., value, difference, proportion) as input and automatically generates a data video by selecting appropriate clips from an animation library. In contrast, Data Animator~\cite{thompson2021data} and CAST~\cite{ge2021cast} enable animation authoring with a graphical user interface (GUI). Users can set up keyframes by importing Data Illustrator projects or selecting graphic components. Then, users can configure animation designs using timing parameters and data mappings between adjacent keyframes. While the aforementioned tools have eased the authoring process, creating a datamation is still not easy for users with little data analysis background. They have to prepare the keyframes or the data facts that are used to synthesize animations. 

To facilitate an easy transition between different forms of visualization views, unit visualization has been introduced~\cite{drucker2015unifying}. It represents data items as units and packs them into various visualization forms that could be flexibly transformed through animated transitions. It has been used in recent studies for authoring animated transitions~\cite{ge2021cast} and data stories~\cite{lu2021automatic}. However, none of these works support a direct authoring of datamations which needs to pay more attention to the analysis pipeline and the underlying data flows. \name also leverages the flexibility of the unit visualization design to animate an analysis pipeline. Moreover, we introduced a set of low-level actions that precisely control the existence, appearance, and layout of the units to help with the datamation generation and authoring.

\subsection{Natural Language Interface for Data Visualization}
To lower the barriers of creating a visualization chart, there are techniques that can automatically generate or recommend visualizations of tabular data input.  Various natural language interfaces (NLIs) for data visualization have been explored both within the research community~\cite{sun2010articulate,gao2015datatone,setlur2016eviza,aurisano2016articulate2,narechania2020nl4dv,hoque2017applying,cox2001multi,wen2005optimization,srinivasan2020interweaving} and industry~\cite{powerbi,tableau}. 

While NLIs provide flexibility in posing data-related questions, inherent characteristics of natural language such as ambiguity and under-specification make precisely understanding user intentions a challenging task. To overcome this obstacle, NLIs are designed to either guide users to provide a more concrete nature language query~\cite{cox2001multi,setlur2016eviza,yu2019flowsense} or untangle ambiguities in the query~\cite{gao2015datatone,wen2005optimization,sun2010articulate} to capture user intent. 

One approach to parsing natural language uses pre-defined grammars.
Flowsense~\cite{yu2019flowsense} and Eviza~\cite{setlur2016eviza} depend on a pre-defined grammar to capture query patterns. 
However, grammar-based methods require tedious manual specifications of query patterns by visualization experts. 
Therefore, many works leverage sophisticated parsing techniques from natural language processing to understand the intention of queries and detect ambiguities present in the queries.
Articulate~\cite{sun2010articulate} allows people to generate visualizations by deriving mappings between tasks and data attributes in user queries, the translation of  imprecise user specification is based on a natural language parser enriched with machine learning algorithms that can make reasoned decisions.  
The most recent work, ncNet~\cite{luo2021natural}, builds a Transformer-based sequence-to-sequence model to translate natural language queries to visualization~specifications. 

Most of the existing NLIs are designed to translate the natural language to one or more static visualizations. How to appropriately generate animation based on natural language has yet to be studied. We address this lack in the literature by introducing a language-driven authoring tool developed to support datamation design and generation.

\section{System Overview}
In this section, we describe the design requirements of the \name system, followed by an introduction of the system architecture and running pipeline.

\subsection{Design Requirements}
The proposed \name system has been designed to meet several real-world requirements for creating an interpretable data analysis pipeline and showing it via data animations. The design objective is to address the challenges discussed in the introduction, whereas the concrete requirements were derived from separate interviews with three domain experts whose expertise was in interactive data analysis and natural language processing. During the interviews, detailed system design requirements were discussed, and a prototype system was demonstrated to the experts to gather feedback. Below, we describe the most critical design requirements that were identified during these discussions, which motivate the design adopted in our work.

\begin{enumerate}

\item[{\bf R1}] {\bf Supporting data queries in natural language.} The system should allow users to input queries about the data in natural language to eliminate the technical barriers of data analysis and datamation authoring.

\item[{\bf R2}] {\bf Uncovering the analysis pipeline for interpretation.} The system should not only be able to find the desired results based on the input query, but more importantly, it needs to resolve the query in the context of the input data to reveal the detailed steps of the underlying analysis pipeline to help interpret the results. 

\item[{\bf R3}] {\bf Creating comprehensible animations.} The system should be able to generate meaningful animations that are easy to follow and understand, illustrating the changes between any of the two succeeding steps in the analysis pipeline without losing focus or increasing cognitive load.

\item[{\bf R4}] {\bf Learning from users' feedback.} To ensure the quality of the generated datamations, the system should allow users to refine the generation results.

\item[{\bf R5}] {\bf Interactive communication and response.} To support a human-machine collaborative design environment, the system should be fast enough to provide a prompt query and feedback mechanism and respond to users without latency. 

\end{enumerate}

\subsection{System Architecture and Running Pipeline}    
To fulfill the above requirements, we introduce the \name system. As shown in Fig.~\ref{fig:pipeline}, it consists of three major modules: (a) the \textit{Preprocessing Module}, (b) the \textit{Data Query Analyzer}, and (c) the \textit{Datamation Editor}. Specifically, when a user uploads a tabular dataset $X$ and input a data query $q$ in natural language ({\bf R1}), the \textit{Preprocessing Module} (Fig.~\ref{fig:pipeline}(a)) parses $X$ and $q$ into a word sequence $q_x$ to facilitate the calculations in the following steps:  
\begin{equation}
{q_x} \leftarrow [w_1, ..., \langle T \rangle, t_{1}, ...,\langle /T \rangle \langle C \rangle, c_{1}, ..., \langle /C \rangle,\langle V \rangle, v_{1}, ...\langle /V \rangle]
\label{eq:question}
\end{equation}
where $w_i$ is a word in $Q$; $t_i$, $c_i$ are the table and column names, respectively; $v_i$ is a query-related data value in $X$ regarding to the column $c_i$. Tokens, i.e., $\langle T \rangle$, $\langle C \rangle$, $\langle V \rangle$, are used to denote the begin and end of different parts. 

In the next step, the \textit{Data Query Analyzer} translates the $q_x$ into a sequence of Question Decomposition Meaning Representation (QDMR) operations, defined by Wolfson \etal~\cite{wolfson2020break}:
\begin{equation}
\label{eq:decompose}
S_{op} = [op_{1}, op_{2}, ..., op_{n}] \leftarrow Decompose ({q_x})
\end{equation}
where each operation $op_i$ indicates a simple calculation (e.g., filter or aggregate) on the input data or the outputs of the previous operation(s). Ideally, executing these operations in order will provide the desired results regarding the data query. In this case, the above operation sequence will indicate a valid analysis pipeline ({\bf R2}).

Finally, each operation in $S_{p}$ will be used to drive an action-oriented dynamic unit visualization to generate a datamation ({\bf R3}). The generated datamation will be displayed in the \textit{Datamation Editor}, in which users can interactively remove an operator, change an operator's order, or add new operators in $S_{op}$ to refine or fix the generation results. The datamation will be updated accordingly in real-time during the editing process ({\bf R5}). Users' modifications will be recorded and fed back into the \textit{Data Query Analyzer} ({\bf R4}), in which a pre-trained deep knowledge editing network (Fig.~\ref{fig:pipeline}(b2)) takes the modified sequence $S'_{p}$ to calculate a set of parameters to calibrate the decomposition model.



\begin{table*}[t]
    \setlength\aboverulesep{0pt}
    \setlength\belowrulesep{0pt}
    \def\arraystretch{1.3}
\caption{The QDMR operations and the corresponding actions designed to drive a dynamic unit visualization.}
\label{tab:operators}
    \vspace{1mm}
\begin{tabular}{p{0.1\linewidth}wc{0.2\linewidth}p{0.3\linewidth}wc{0.075\linewidth}wc{0.075\linewidth}wc{0.075\linewidth}}
\toprule

\multicolumn{3}{c|}{\textbf{(a) QDMR OPERATIONS}}           & \multicolumn{3}{c}{\textbf{(b) UNITVIS ACTIONS}} \\\midrule
\textbf{Operator} &
\multicolumn{1}{c}{\textbf{Arguments}}    &
 \multicolumn{1}{c|}{\textbf{Description}}  &
 \multicolumn{1}{c}{\textbf{Data}} &
 \multicolumn{1}{c}{\textbf{Visual}}  &
 \multicolumn{1}{c}{\textbf{Annotation}} \\\midrule
SELECT &  \begin{tabular}[c]{@{}c@{}}
table/column, \\condition
\end{tabular}   &
\multicolumn{1}{l|}{\begin{tabular}[c]{@{}l@{}}
Select data records from a data source (i.e.,\\table/column) for the given \textit{condition}
 \end{tabular}}
& select & layout & /
\\\hline

PROJECT & records, attribute     
& \multicolumn{1}{l|}{Retrieve the \textit{attribute} values from the \textit{records}}  & / & 
\begin{tabular}[c]{@{}c@{}}
size [numerical] \\ color [categorical] \\x-axis [temporal]
\end{tabular}

& /
\\ \hline
COMPARATIVE  & 
\begin{tabular}[c]{@{}c@{}}
records, attribute, \\condition
\end{tabular}
& 
\multicolumn{1}{l|}{
\begin{tabular}[c]{@{}l@{}}
Filter out the \textit{records} whose \textit{attribute} value\\
 satisfies ($=, \neq, >, <, \ge, \le$) a condition  \end{tabular} }
 & filter & / & highlight,hide
 \\\hline

SUPERLATIVE  & 
\begin{tabular}[c]{@{}c@{}}
records, attribute, \\superlative condition  
\end{tabular}
& \multicolumn{1}{l|}{\begin{tabular}[c]{@{}l@{}}Find a record from the \textit{records} whose \\ \textit{attribute} has the maximum/minimum value\end{tabular}}  & filter & / & highlight,hide                      
\\ \hline
AGGREGATE    & records, agg methods                  &  
\multicolumn{1}{l|}{\begin{tabular}[c]{@{}l@{}} Compute the \textit{max/min/sum/count/avg}\\ value of the records
\end{tabular}} & aggregate & / & annotate                             
\\ \hline
GROUP        & 
\begin{tabular}[c]{@{}c@{}}
records, attribute \\ agg methods 
\end{tabular}  
& \multicolumn{1}{l|}{\begin{tabular}[c]{@{}l@{}}
Group \textit{records} by the \textit{attribute} and \\ compute the \textit{max/min/sum/count/avg} \\ value of each group
\end{tabular}}  & / & 
\begin{tabular}[c]{@{}c@{}}
x-axis[temporal]\\
y-axis[categorical]
\end{tabular}
& annotate
\\ \hline
UNION        & $records_1$, $records_2$  &  
\multicolumn{1}{l|}{\begin{tabular}[c]{@{}l@{}} Combine \textit{$records_1$} and \textit{$records_2$}
\end{tabular}}  
& union & / & / 
\\ \hline
DISCARD & $records_1$, $records_2$   & \multicolumn{1}{l|}{\begin{tabular}[c]{@{}l@{}} Find the instances in \textit{$records_1$} but \\not in \textit{$records_2$}.
\end{tabular}}   & filter & / & hide         \\ \hline
INTERSECTION      & $records_1$, $records_2$ & \multicolumn{1}{l|}{\begin{tabular}[c]{@{}l@{}} Find the instances belonging to both \\ \textit{$records_1$} and \textit{$records_2$}
\end{tabular}}   & intersection & / & hide   
\\ \hline
SORT         & 
\begin{tabular}[c]{@{}c@{}}
records, attribute,\\ \textit{asc/desc}
\end{tabular} 
& \multicolumn{1}{l|}{\begin{tabular}[c]{@{}l@{}}Sort the \textit{records} by \textit{attribute} in a \\\textit{asc/desc} order
\end{tabular}}   & sort & / & /   
 \\\bottomrule          
\end{tabular}
\vspace{-1.5em}
\end{table*}

\section{Data Query Analyzer}

\begin{figure}[!tbh]
\centering 
\includegraphics[width=0.98\columnwidth]{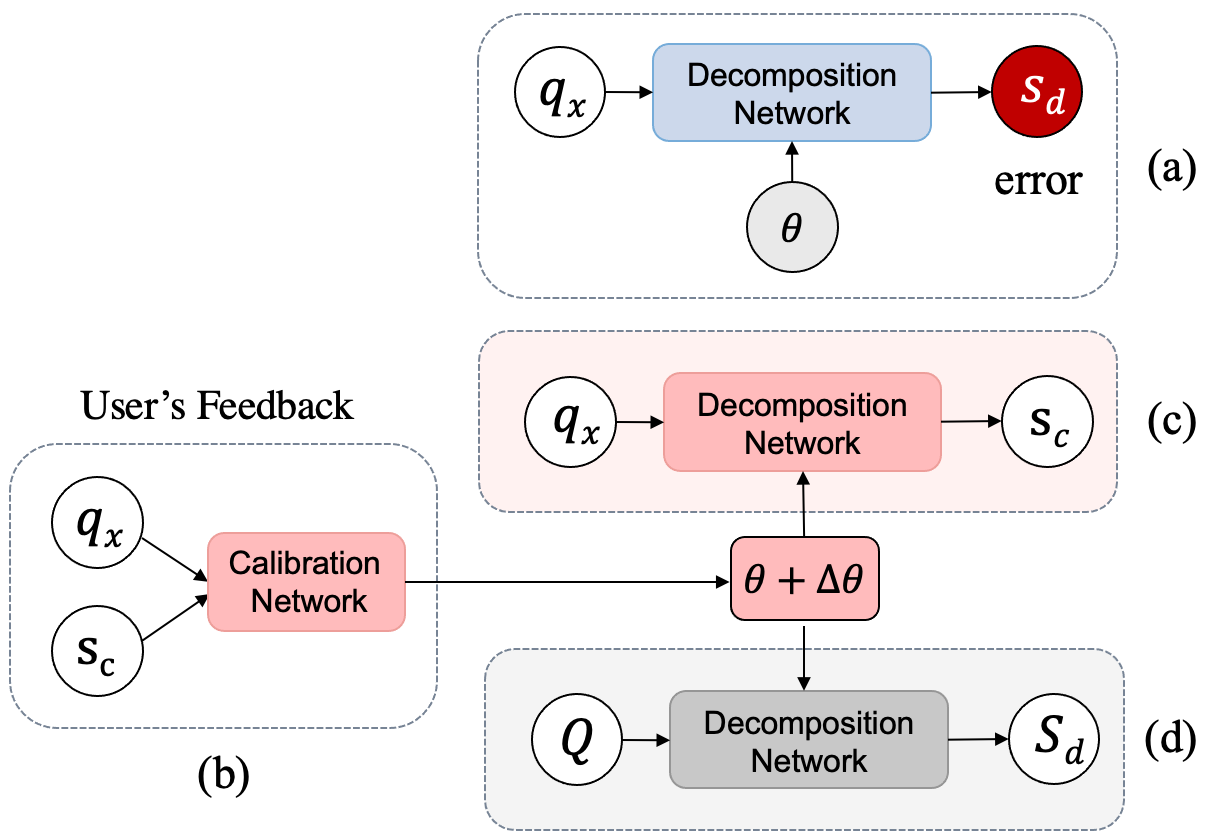}
\caption{Data query decomposition with online knowledge calibration.}
\vspace{-1.5em}
\label{fig:calibrate}
\end{figure}

The \textit{Data Query Analyzer} is designed to resolve an input natural language data query in the context of a given dataset and decompose it into a sequence of data operations in the form of QDMR~\cite{wolfson2020break}. When a decomposition error occurs, it will modify the result based on users' feedback. Specifically, as shown in Fig.~\ref{fig:calibrate}, the design of the \textit{Data Query Analyzer} consists of two parts: the decomposition network ($\mathcal{D}$) and the calibration network ($\mathcal{C}$). When $\mathcal{D}$ wrongly resolves a data query $q_x$ into a problematic operation sequence $s_d$ (Fig.~\ref{fig:calibrate}(a)), the user can modify it to provide a calibration $s_c$. $\mathcal{C}$ takes $(q_x, s_c)$ as the input and produces a parameter calibration $\Delta \theta$ for $\mathcal{D}$ (Fig.~\ref{fig:calibrate}(b)). By adding $\Delta \theta$ to $\mathcal{D}$'s parameter $\theta$, the decomposition network will be able to generate correct results given $q_x$ or the questions similar to $q_x$ (Fig.~\ref{fig:calibrate}(c)) without affecting the decomposition results of other questions (Fig.~\ref{fig:calibrate}(d)). Next, we will briefly review the definition of QDMR operations and then introduce the technical details of the proposed decomposition network and the corresponding feedback mechanism. 

\subsection{Question Decomposition Meaning Representation}
\label{sec:qdmr}

QDMR is an approach to representing a question by a sequence of operations that can be executed to answer the question~\cite{wolfson2020break}. Each operation is responsible for querying the source data or analyzing the outputs of the previous operation(s). It is a data-independent representation that can be applied to many NLP benchmarks. Wolfson~\etal~\cite{wolfson2020break} released Break, a question decomposition dataset of 83,978 questions over ten NLP benchmarks, such as data-related questions from Spider~\cite{yu-etal-2018-spider}, document-related questions from HotPotQA~\cite{yang-etal-2018-hotpotqa}, and image-related questions from CLEVR~\cite{johnson2017clevr}. In our work, we focus on ten types of operations employed to represent the data-related questions in Spider~\cite{yu-etal-2018-spider}, and use them to drive the datamation generation. The formalization and detailed explanation of each operation are provided in Table~\ref{tab:operators}(a).

\begin{figure}[!tbh]
\centering 
\includegraphics[width=0.95\columnwidth]{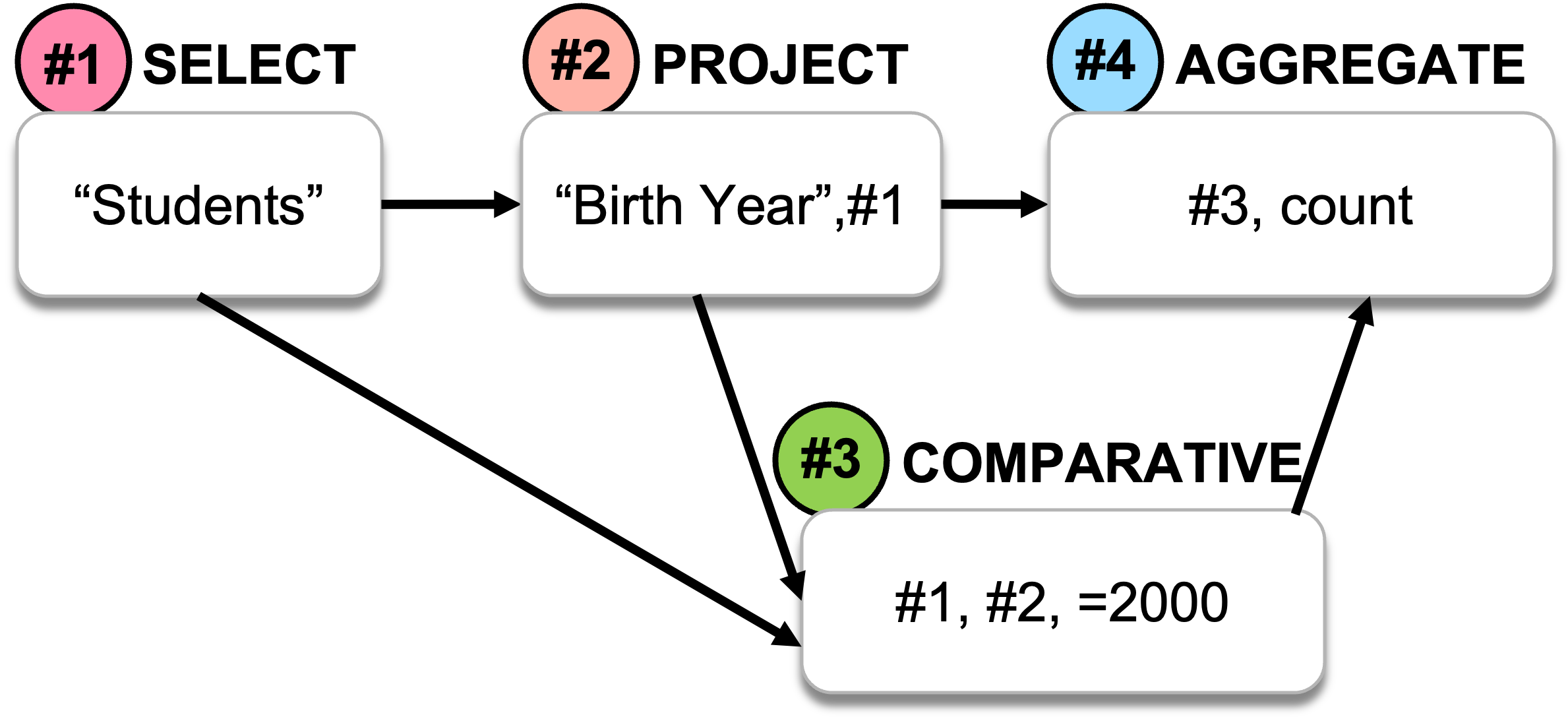}
\caption{QDMR operations chain to answer the question: ``how many students were born in 2000?''}
\vspace{-1.5em}
\label{fig:qdmr}
\end{figure}

Fig.~\ref{fig:qdmr} demonstrates an example of using QDMR to resolve the question: \textit{how many students were born in 2000?} A sequence of four operations is used to answer this question:
\begin{itemize}
    \item Step 1: \texttt{SELECT} (``Students"). \texttt{SELECT} is an operation used to retrieve records from the data based on certain conditions. It is similar to the ``select'' statement in the structured query language (SQL). The argument \textit{``Students''} corresponds to a table name in the dataset. In step 1, the intent of the operation is to select all the student records from the data for further exploration.
    
    \item Step 2: \texttt{PROJECT} (``Birth Year", \#1).  \texttt{PROJECT} is an operation that retrieves a certain attribute from the source/input records. The argument \textit{``Birth Year''} specifies the attribute to be retrieved, which is usually a data column; \textit{\#1} corresponds to the output of step 1. In step 2, the intent of the operation is to retrieve the year of birth of previously selected students.
    
    \item Step 3: \texttt{COMPARATIVE}(\#1, \#2, ``$=$2000"). The \texttt{COMPARATIVE} operation uses a comparative condition to filter records based on a specified attribute. The argument \textit{\#1} denotes the records that need to be filtered; the second argument \textit{\#2} is the attribute based on which the data will be filtered (here, the birth year of students). The last argument, \textit{``$=$2000''}, is the comparative condition based on which the data will be filtered. Step 3 aims to find out the students born in 2000.
    
    \item Step 4: \texttt{AGGREGATE}(\#3, count). \texttt{AGGREGATE} computes the aggregated value of the records. QDMR supports six frequently used aggregation methods: count, max, min, sum, avg, and median. The first argument refers to the data records to be aggregated. The second argument indicates the aggregation method. Step 4 calculates the number of students in the filtered records.

\end{itemize}

QDMR has been used in various tasks, such as open domain QA~\cite{wolfson2020break}, natural language for executable database queries~\cite{saparina-osokin-2021-sparqling}, and contrast set generation~\cite{geva2021break}. In our work, we use QDMR as a structured intermediate representation of questions, connecting to datamation generation. Each type of QDMR operation maps to a series of low-level actions that drive a captioned unit visualization for the animation. After resolving a question into a sequence of QDMR operations, these operations can be used to generate a datamation to visualize the changes of the intermediate analysis results through animated transitions.



\subsection{Decomposition Network ($\mathcal{D}$)}
\label{sec:decomposition}
We adopt and fine-tune a pre-trained natural language model T5~\cite{wolf2019huggingface} that was developed based on the Transformer architecture~\cite{vaswani2017attention} to translate the input data query (and the corresponding data scheme) to a sequence of QDMR operations. T5 is used due to its many advantages shown in a wide range of translation-related tasks, such as natural language translation~\cite{vaswani2018tensor2tensor}, text2sql~\cite{rat-sql}, and text summarization~\cite{liu2019text}. We fine-tune the model based on the aforementioned QDMR data corpus. To make it simple, we present the conceptual calculation steps in the context of our problem to show a brief idea about the model but leave the mathematical details to Vaswani~\etal~\cite{vaswani2017attention}.

Generally, a Transformer follows the encoder-decoder architecture. Given a preprocessed data query $q_x$, the model first embeds $q_x$ into a vector representation $v$. Then, it encodes $v$ into a latent vector $h_e$ that captures the semantics of the data query $q$ and the corresponding tabular data $X$:
\begin{equation}
h_e = encode(v), ~~v = embed(q_x)
\end{equation}
Later, the decoder takes $h_e$ to compute a decoded latent vector $h_d$ and then uses it to compute the output probabilities of the tokens in the vocabulary given by the training samples via a softmax layer:
\begin{equation}
h_d = decode(h_e)
\end{equation}
\begin{equation}
P_{voc} = softmax(W h_d^\top)
\end{equation}
where $W$ is a weight matrix to be trained. In each round, the token in the vocabulary having the highest probability is chosen as the output of the model. In this way, the model generates a QDMR operation sequence token by token in the following form:
\begin{equation}
\label{eq:decompose}
[op_1\{\mathtt{t}_{11},\ldots, \mathtt{t}_{1j}\}, \ldots,op_n\{\mathtt{t}_{n1}, \ldots, \mathtt{t}_{nk}\}] \leftarrow decode (h_e)
\end{equation}
where $\mathtt{t}_{ij}$ indicates the $j$-th token of the $i$-th operator.

\paragraph{\bf Loss Function}
To encourage the output of the decomposition network as identical as possible with the target analysis pipeline in our training corpus, the model is trained by using cross-entropy loss between generated operation sequence and ground-truth:
\begin{equation}
L =-\sum_{i=1}^{n} \log p(t_{i} \mid t_{1}, \ldots, t_{i-1}, q_x)
\label{eqn:decompositionloss}
\end{equation}
where $t_i$ is the current reference token in the target pipeline. Given the previous tokens $t_{1}, \ldots, t_{i-1}$ and the input data query $q_x$, this loss function tends to maximize the probability of the reference token $t_i$ as the prediction in the current step.

\paragraph{\bf Training Corpus} We adopt the dataset introduced in~\cite{saparina-osokin-2021-sparqling} to train our decomposition network. It was generated by manually annotating the Spider dataset~\cite{yu-etal-2018-spider} based on the QDMR operations. In particular, it contains 7,423 natural language data queries about a number of databases introduced in Spider. Each query corresponds to a sequence of manually annotated QDMR operations together with attributes such as the table and column names. 

\paragraph{\bf Implementation} Our decomposition network was implemented based on PyTorch~\cite{paszke2019pytorch} and fine-tuned via 20 epochs with the gradient accumulation step as 16. The batch size was set to 8. The Adam~\cite{diederik2014adam} optimizer was used and the learning rate was set to 2e-4. The overall training procedure spent around 2 hours on an Ubuntu server with an NVIDIA V100 GPU.

\subsection{Calibration Network ($\mathcal{C}$)}
\label{section:calibrationnetwork}
Although the above model is effective, it cannot always guarantee a correct decomposition result. Sometimes, disordered analysis pipelines or pipelines with missing operations are produced. It makes sense to let users interactively fix these issues and use their feedback to polish the decomposition model. However, it usually takes a long period of time to accumulate enough feedback to re-train or fine-tune the deep learning model, thus making quick responses difficult. To address this issue, as shown in Fig.~\ref{fig:calibrate}, instead of using the feedback as the new training samples, we introduce an efficient online post-hoc knowledge calibration mechanism. Here, we compute a parameter calibration (denoted as $\Delta \theta$) based on users' feedback to modify the parameters (i.e., the knowledge learned from the training samples, denoted as $\theta$) of the original decomposition network to help produce the desired result as suggested in the user's feedback.

\begin{figure}[tb]
\centering 
\includegraphics[width=0.45\textwidth]{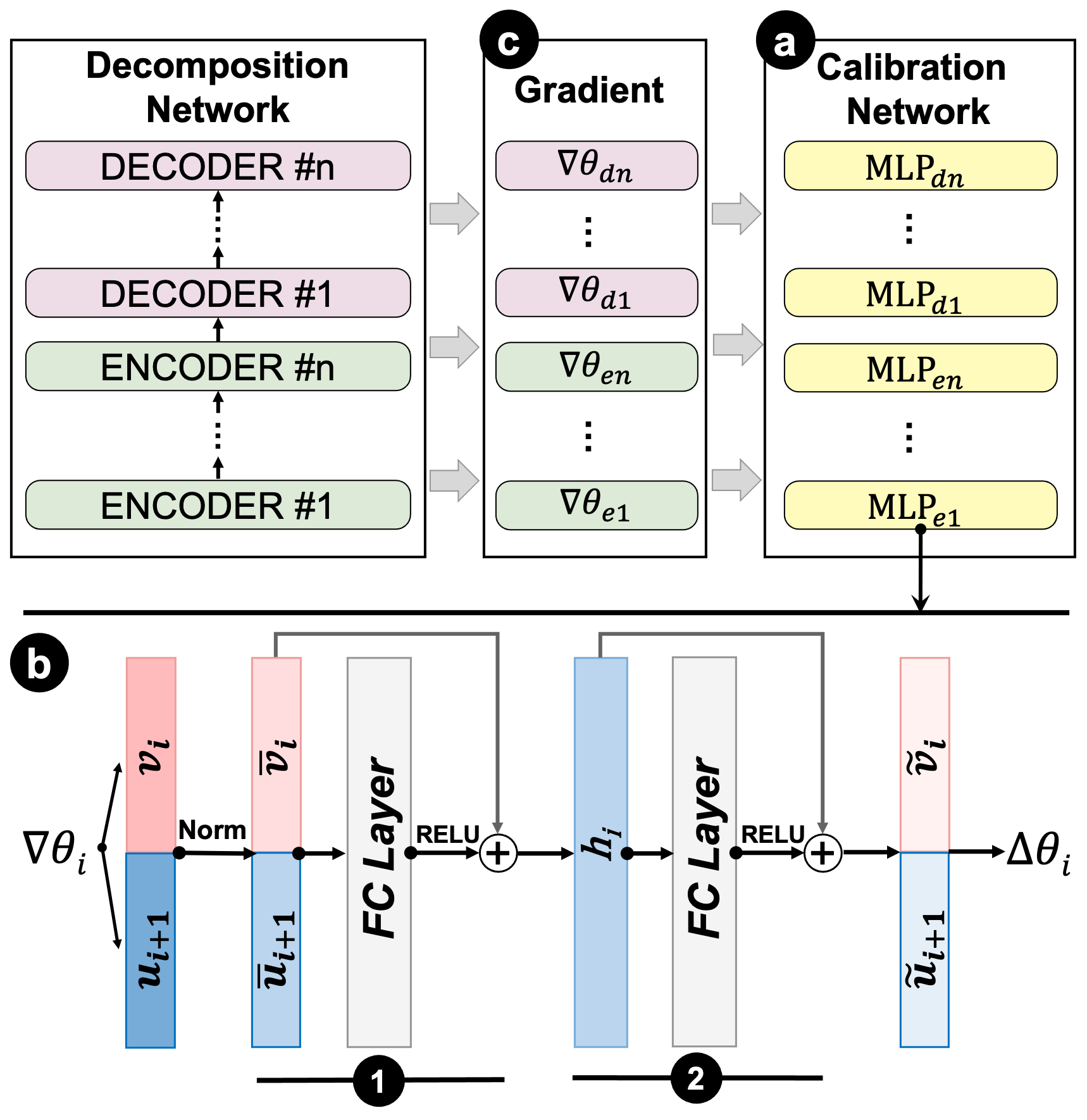}
\vspace{-1.0em}
\caption{The architecture of the calibration network.}
\label{fig:editnetwork}
\end{figure}



A deep knowledge calibration network (Fig.~\ref{fig:editnetwork}(a), denoted as $\mathcal{C}$) is designed to implement the above idea. It is a collection of auxiliary multi-layer perceptrons (MLPs). We choose to use MLP for building the network due to many of its advantages such as excellent fault tolerance and strong adaptive and self-learning features. Each MLP is responsible for calibrating the parameters of a corresponding layer in the decomposition network toward the direction of making the model output the desired operation sequence $s_c$ given by the users' feedback regarding the data query $q_x$. We implement the MLP with a single hidden layer and adopt skip connection~\cite{he2016deep} to improve the performance and convergence of the MLP (Fig.~\ref{fig:editnetwork}(b)). The input of each MLP is the loss gradient ($\nabla_{\theta_{i}}L(q_x,s_c)$, Fig.~\ref{fig:editnetwork}(c)) of the decomposition network in the $i$-th layer calculated when fine-tuning the decomposition network based on user's feedback $s$ under the parameter setting of $\theta_i$. The outputs of the MLP are the parameters $\Delta \theta_{i}$ that calibrate $\theta_{i}$ to make the decomposition network achieve a result that matches $s_c$. In particular, the loss gradient in the $i$-th layer is calculated based on the chain-rule during a back-propagation process when fine-tuning the decomposition network based on $s_c$. As the feedback $s_c$ is processed token by token, the gradient is thus computed token by token as well. We compute an averaged gradient that could be used as the input to the aforementioned MLP:  
\begin{equation}
\nabla_{\theta_{i}}L(q_x,s_c) = \frac{1}{n}\sum_{j=1}^n \nabla_{\theta_{i}}L(q_x,j) =  \frac{1}{n}\sum_{j=1}^n u_{(i+1)j} v_{ij}
\label{eqn:gradient0}
\end{equation}
where $n$ is the total number of tokens in $s$; $u_{(i+1)j}$  and $v_{ij}$, respectively, indicate the loss gradient in the last layer and the hidden vector in the currently layer corresponding to the $j$-th token in $s_c$. Equation~(\ref{eqn:gradient0}) is derived based on the chain rule of the back-propagation process. Details can be found in the book by Goodfellow \etal~\cite{Goodfellow-et-al-2016} (Section~6.5). 

However, the above gradient is a high-dimensional vector. Therefore, it requires even more parameters to directly train a network that maps such a gradient to a parameter calibration. It could be millions of parameters to tune when calibrating a large decomposition model like T5, making it very difficult to converge. To address this issue, we leverage the gradient decomposition strategy~\cite{mitchell2021fast} to reduce the number of parameters used in the calibration network. The idea is to directly map each gradient corresponding to a token $j$ in $s_c$ into a parameter calibration $\Delta \theta_{ij}$ independently via an MLP, and then use the average as the overall calibration $\Delta \theta_{i}$:
\begin{equation}
\Delta \theta_{i} = \frac{1}{n}
\sum_{j=1}^n \Delta \theta_{ij} = \frac{1}{n}
\sum_{j=1}^n \widetilde{u}_{(i+1)j} \widetilde{v}_{ij}
\label{eqn:gradient}
\end{equation}
where $\widetilde{u}_{(i+1)j}$ and $\widetilde{v}_{ij}$ are the output of an MLP that are derived based on $u_{(i+1)j}$ and $v_{ij}$ as follows:
\begin{align} 
&z_{ij} = concat(norm({u}_{(i+1)j}) ,norm({v}_{ij})) \label{eqn:zconcat}\\
&h_{ij} = s_{i} \odot (z_{ij} + \sigma(U_1V_1z_{ij})) + o_{i} \label{eqn:mlp1stblock}\\
&\widetilde{z_{ij}} = s'_{i} \odot (h_{ij} + \sigma(U_2V_2h_{ij})) + o'_{i} \label{eqn:mlp2ndblock}\\
&\widetilde{u}_{(i+1)j},~\widetilde{v}_{ij} = split(\widetilde{z_{ij}})
\label{eqn:zsplit}
\end{align}
where Eq.~(\ref{eqn:mlp1stblock}) and Eq.~(\ref{eqn:mlp2ndblock}), respectively, indicate the computation of the two consecutive blocks in MLP as shown in Fig.~\ref{fig:editnetwork}(b1,b2). In particular, the first block (Eq.~(\ref{eqn:mlp1stblock})) takes a vector $z_{ij}$ that concats the normalized $u_{(i+1)j}$ and $v_{ij}$ (Eq.~(\ref{eqn:zconcat}) as the input followed by a fully connected layer whose weight matrix is factorized into $U_1$ and $V_1$, which will be gradually transformed into the final output. Here, we use ReLU (denoted as $\sigma(\cdot)$) as the activation function whose output is directly added with the input $z_{ij}$ through a skip connection in purpose of mitigating the degradation issues during the training process. $s_{i}$ and $o_{i}$ are the trainable scale and offset used to regularize the output of the first block before passing it to the next step. The second block takes the previous output $h_i$ to perform a similar computation as shown in Fig.~\ref{fig:editnetwork}(b2) and described in Eq.~(\ref{eqn:mlp2ndblock}). The final result $\widetilde{z_{ij}}$ is split into the desired $\widetilde{u}_{(i+1)j}$ and $\widetilde{v}_{ij}$ (Eq.~(\ref{eqn:zsplit})).

Finally, we update the parameter $\theta_{i}$ in the $i$-th layer of the decomposition network by $\theta'_{i}$ as follows:
\begin{align} 
\theta'_{i} = \theta_{i} + \Delta \theta_{i}
\end{align}

\paragraph{\bf Loss Function} The above calibration network is trained one feedback $(q_k, s_{ck})$ a time based on a set of training samples consisting of data queries $Q_x = \{q_1, \cdots, q_m\}$. Their decomposition results $S_d = \{s_{d1}, \cdots, s_{dm}\}$ are generated before calibrating the decomposition network, and the desired results $S_c = \{s_{c1}, \cdots, s_{cm}\}$ are given by users'~feedback. In each round of training, the following loss function is~minimized:
\begin{align} 
L = \alpha L_r + L_p
\label{eq:loss}
\end{align}
where $L_r$ is the cross-entropy loss that estimates the similarity between the decomposition result generated after calibrating the decomposition network using $\theta$' and the desired results indicated in a user's feedback~$s_c \in~S_c$:
\begin{equation}
L_r =-\sum_{i=1}^{n} \log p_{(\theta')}(t_{i} \mid t_{1}, \ldots, t_{i-1}, q_k)
\end{equation}
where $p_{(\theta')}(\cdot)$ estimates the probability of generating a token $t_i$ that belongs to a desired decomposition result in $s_c \in S_c$, given all the previous tokens $t_{1}, \ldots, t_{i-1}$ and the data query $q_k$, after calibrating the decomposition network based on $\theta'$.

In Eq.~(\ref{eq:loss}), $L_p$ is the Kullback-Leibler divergence that estimates the similarity of the output distributions before and after calibrating the decomposition network $\mathcal{D}$ regarding to the remaining data queries  $Q'_x = Q_x - \{q_k\}$:
\begin{equation} 
L_p = KL(\mathcal{D}(Q'_x|\theta), \mathcal{D}(Q'_x|\theta'))
\end{equation}
Intuitively, minimizing $L_r$ ensures that the feedback will be accepted by the decomposition network and minimizing $L_p$ will prevent our calibration from affecting the decomposition results beyond the feedback.

\paragraph{\bf Training Corpus} We prepared a corpus with 12k data samples to train the calibration network. Each sample is a triplet that consists of a data query $q_x$, the corresponding problematic decomposition result $s_d$, and a correction $s_c$, denoted as $(q_x, s_d, s_c)$. To collect these data samples, we deliberately trained a decomposition model with a low accuracy based on T5~\cite{wolf2019huggingface} by only using a small subset of training samples randomly selected from the aforementioned corpus. We tested the entire datasets on the model and selected the incorrect outputs $s_d$ together with the corresponding data query $q_x$ and ground truth $s_c$ into our calibration data set. We iteratively performed the above training and testing process by using a different subset of samples to train the model every time. We enriched the data queries by translating the original English queries into Chinese and then translating it back to English via Google translate. As a result, a number of 12,227 unique training samples were collected, which covered two types of decomposition errors: missing operations (3340 samples) and disordered sequences (4693 samples).

\paragraph{\bf Implementation}
The calibration network was also implemented in PyTorch. We chose Adam~\cite{diederik2014adam} as the optimizer and set the learning rate as 1e-4. The overall training procedure spent around 16 hours on an Ubuntu server with one NVIDIA V100 GPU.

\begin{figure}[tb]
\setlength{\abovecaptionskip}{10pt}
\centering 
\includegraphics[width=\columnwidth]{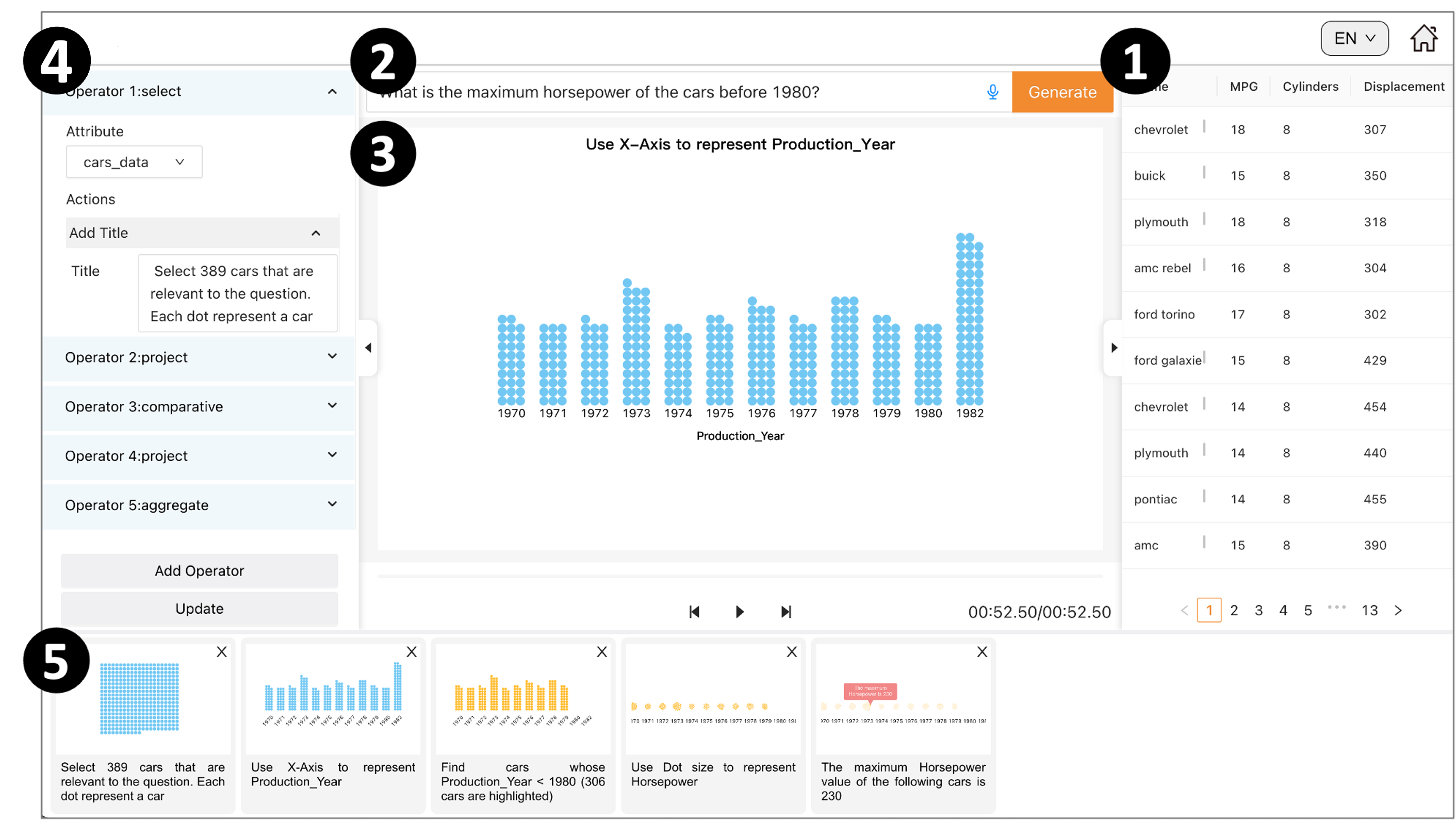}
\vspace{-1.5em}
\caption{The user interface of the datamation editor.}
\vspace{-2.0em}
\label{fig:userinterface}
\end{figure}

\section{Datamation Editor and Animation Generation}
In this section, we first describe the design of the datamation editor and the corresponding interactions. After that, we introduce how the QDMR operation sequences are visually represented by an action-orientated unit visualization to generate datamations.

\subsection{User Interface and Interactions}
The \textit{datamation editor} is designed to translate the QDMR operation sequence into a datamation and help users refine it, and eventually acquire a datamation with correct analysis results and coherent animations. In particular, when a user uploads a spreadsheet into the system, the raw data is displayed in the data panel~(Fig.~\ref{fig:userinterface}-1), which allows users to easily access the data during the authoring process. The user can preview the data and communicate an analysis task of interest by typing a data query into the input box~(Fig.~\ref{fig:userinterface}-2). By resolving the query input, the underlying system will generate an operation sequence to create an initial datamation. The datamation is displayed in the center of the interface~(Fig.~\ref{fig:userinterface}-3) with playback buttons at the bottom to control the datamation. Meanwhile, the key-frames of datamation, which correspond to QDMR operations, are arranged and visualized (Fig.~\ref{fig:userinterface}-5) to illustrate the intermediate analysis result of each operation. Users can drag to rearrange their orders when necessary. Once modified, the order of the operations shown
in the configuration panel~(Fig.~\ref{fig:userinterface}-4) will be updated accordingly. Through the configuration panel, users can edit each of the QDMR operations or add new operations in the analysis pipeline. In particular, the users can modify an operation's parameter and the corresponding actions used to update the unit visualization to fully control the design of the datamation. Finally, by clicking the update button, users' modifications can be fed back into the calibration network to update the decomposition model.

\subsection{Action-Oriented Unit Visualization}
We developed an action-oriented dynamic unit visualization to visualize and animate an analysis pipeline represented in the form of a sequence of decomposed QDMR operations as a datamation. The proposed unit visualization represents each data item via a mark, known as a unit, whose existence, appearance, and layout are precisely controlled by a collection of low-level actions. In this way, by translating each QDMR operation into a series of these actions, we are able to control the visualization to dynamically represent the detailed analysis steps via the animated transitions between succeeding unit visualization views that represent the intermediate analysis results. Our design was implemented based on \textbf{D3.js}~\cite{bostock2011d3}. 

In particular, the following three types of low-level actions are developed, which are respectively responsible for processing the underlying data~(\textit{data actions}), manipulating the visual encoding methods~(\textit{visualization actions}), or adding annotations on units~(\textit{annotation actions}): 

\paragraph{\bf Data Actions} The data actions are designed to update the data items represented in the unit visualization via the following approaches: (1)~\textbf{\textit{select}} a set of qualified data items; or (2)~\textbf{\textit{filter}} the current data items based on a condition; or (3)~\textbf{\textit{union}} or \textbf{\textit{intersect}} two subgoups of data items shown in the view; or (4)~\textbf{\textit{aggregate}} the data items to calculate a statistic measure such as minimum, mean, and sum; or (5)~\textbf{\textit{sort}} the data items in order. When these actions are performed, the visualization layout will be updated accordingly to add, remove, merge, or rearrange the units shown in the visualization.

\begin{figure}[tb]
\setlength{\abovecaptionskip}{10pt}
\centering 
\includegraphics[width=\columnwidth]{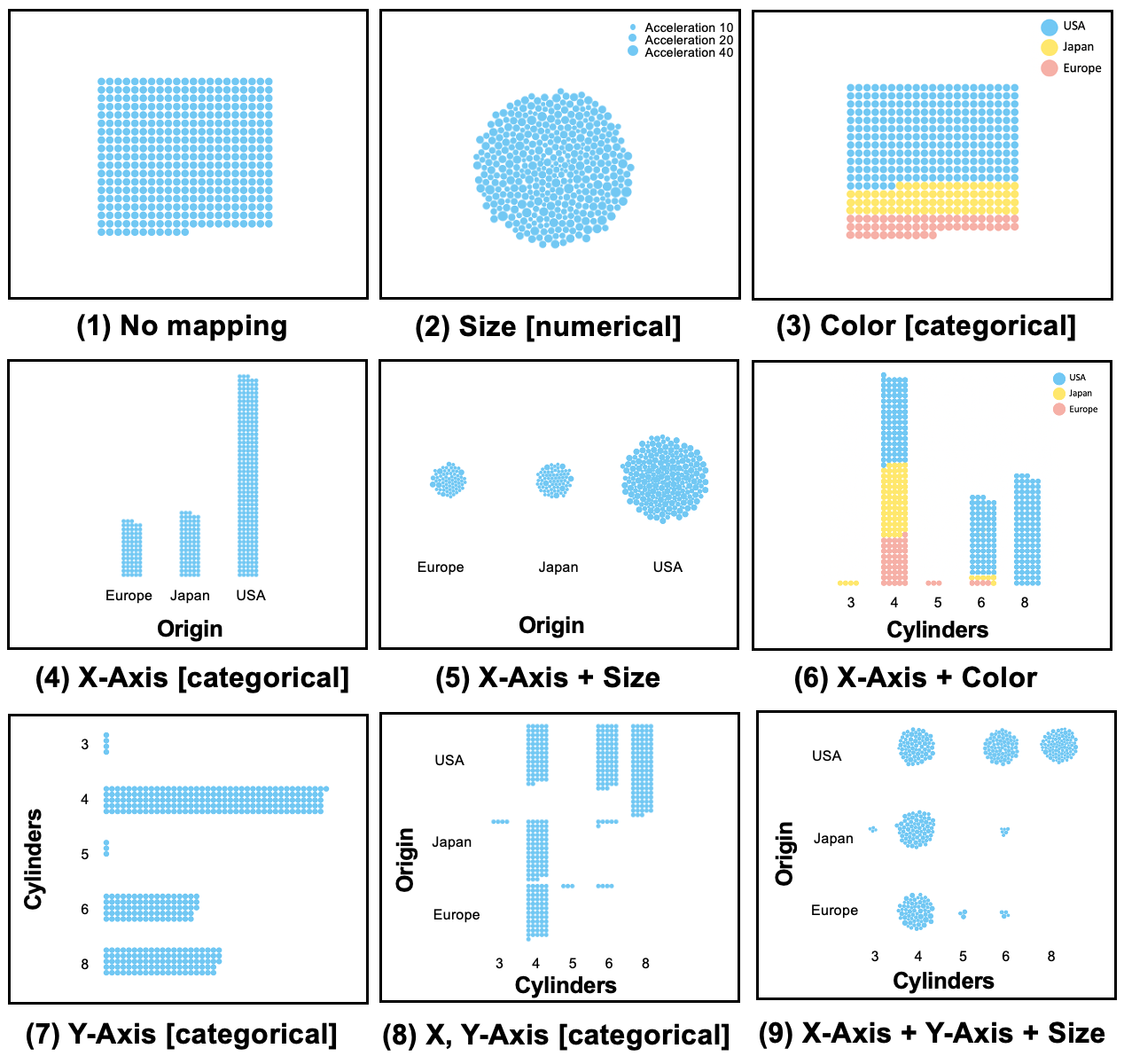}
\vspace{-2em}
\caption{The appearance of the unit visualization is controlled by a set of visual mapping actions.}
\label{fig:unitvisexample}
\vspace{-1.5em}
\end{figure}

\paragraph{\bf Visualization Actions} In our design, the visual appearance (position, color, and size) of each unit in the visualization is determined by the data mappings through four independent visual channels, i.e., x-axis, y-axis, unit size, and unit color designed for mapping different types of data attributes. The visualization actions were designed to control the data mapping of these channels. In particular, \textbf{\textit{x}} and \textbf{\textit{y}} actions are used for mapping numerical, categorical, or temporal data attributes via \{x\}-axis and \{y\}-axis, resulting either in a scatter plot view for representing data with continuous numerical attributes or a group view for the data with discrete categorical attributes (Fig.~\ref{fig:unitvisexample}(4-9)). The \textbf{\textit{size}} action was designed to encode a numerical data attribute by the size of a unit. By default, the size of each unit is mapped to unit 1 and the units in the visualization are packed in a square form to facilitate counting (Fig.~\ref{fig:unitvisexample}(1)). When the size channel is used to map a numerical attribute, the size of each unit is proportional to the corresponding attribute value, and the layout of the units will be changed to circle packing to avoid the overlaps of the units with different sizes (Fig.~\ref{fig:unitvisexample}(2)). Finally, the \textbf{\textit{color}} action is used to map a categorical data attribute by the filling colors(hue) of the units (Fig.~\ref{fig:unitvisexample}(3,6)).  Combining these actions will provide flexible data mapping strategies, resulting in a variety of unit visualization forms with different layouts as shown in Fig.~\ref{fig:unitvisexample}.

\paragraph{\bf Annotation Actions} These actions were designed to \textbf{\textit{highlight}} a focal unit by changing its filling color, or \textbf{\textit{delight}} the non-focal units by making it invisible, i.e., hidden in the view, or \textbf{\textit{annotate}} a unit or a group of units by adding a textual tooltip that shows the specific data attributes of the unit or group. The annotation actions will not affect the change of the layout. Their effects will just be shown through a fade-in or fade-out animation.


\begin{figure*}[t]
\setlength{\abovecaptionskip}{10pt}
\centering
\includegraphics[width=0.95\textwidth]{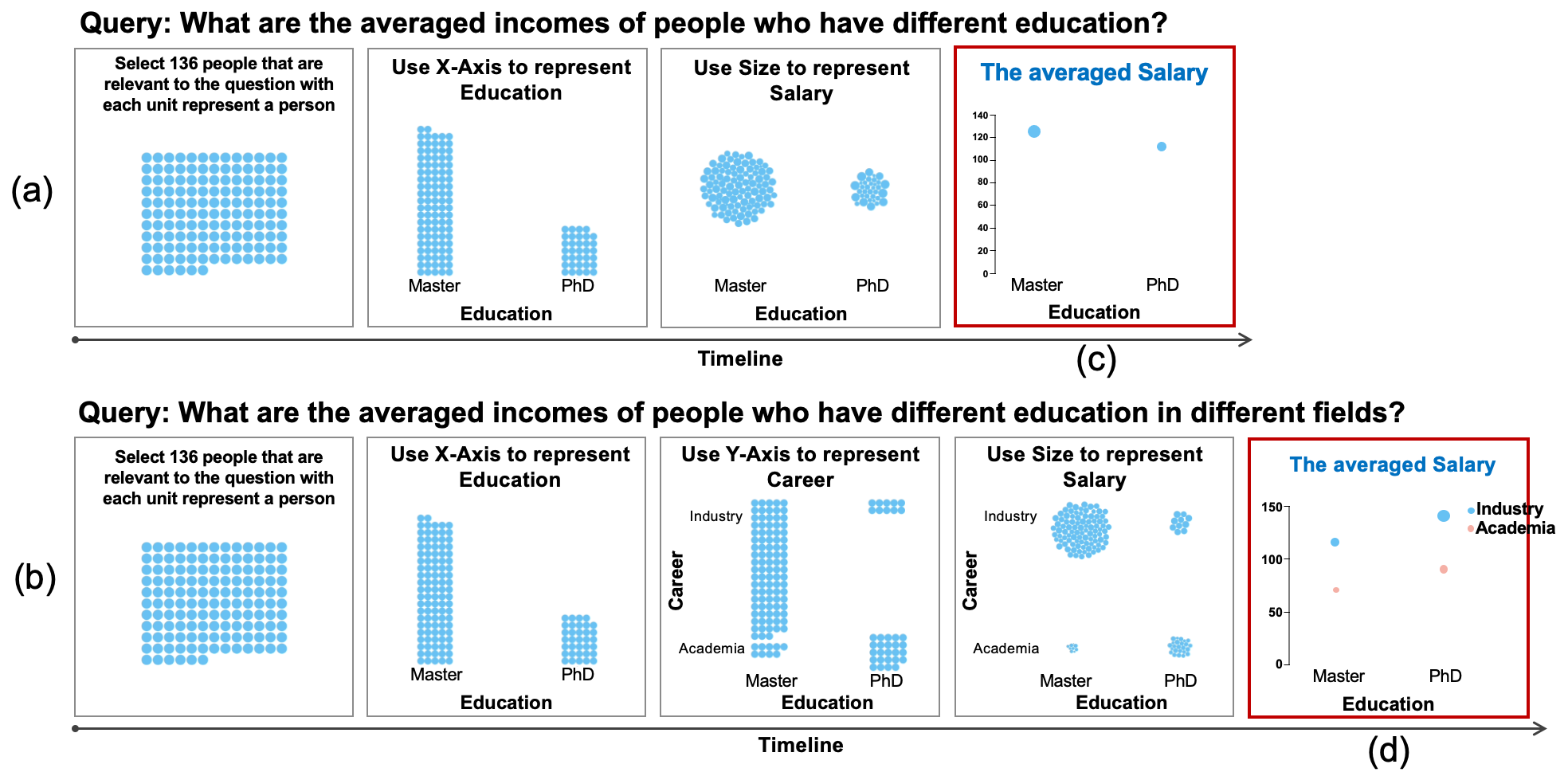}
\vspace{-1.2em}
\caption{In the user study, the participants were asked to resolve a paradox shown in this figure: (a) people with a higher education have lower average income; (b) however, when we take the work field into consideration, the contrary is true.}
\label{fig:userstudy}
\vspace{-1.5em}
\end{figure*}

\subsection{Datamation Generation}
With the above actions in mind, we create a smooth and meaningful datamation via the following three steps: 

\paragraph{\bf Reordering}
The major changes in an analysis process are related to the data produced
by the intermediate analysis steps, thus it is important to first sort the QDMR operations in order to provide a fluent data flow so that the output of the current operation is the input of the next operation. Although in most cases, the decomposition model will produce QDMR operations in the right order, the system will still double-check the decomposition results and reorder the operations to ensure a smooth data flow when necessary. When the automatic reordering fails, errors will be reported for users to fix. Such a failure is usually caused by a problematic decomposition result that contains a wrong type of QDMR operation.

\paragraph{\bf Translation} In the second step, we translate each operation into a sequence of the aforementioned actions based on the rules described in Table~\ref{tab:operators}. These actions are also arranged in order to guarantee a smooth transition. In particular, we first perform the data actions to update the data items shown on the view and then gradually modify their encoding methods and add annotations via the visualization and annotation actions in the follow-up steps. 

\paragraph{\bf Captioning}
Finally, to help communicate the operation's meaning and effects, we generate a caption for each QDMR operation based on a pre-defined template by considering its corresponding low-level actions. The caption describes the change in the data and the visual mappings. It is displayed on top of each view and updated gradually during the animation process. For example, the \texttt{PROJECT} operation can be described as ``use size/color/x-axis to present/encode [attribute]" and the \texttt{AGGREGATE} operation is described as ``the maximum/minimum/average/total value of [a numerical attribute] of the following [units] is [a numeric value]".

Now, let us take the following operation sequence introduced in Section~\ref{sec:qdmr} as an example to illustrate how to generate a datamation based on an input QDMR operation sequence:
\[
[\texttt{SELECT}][\texttt{PROJECT}][\texttt{COMPARATIVE}][\texttt{AGGREGATE}] 
\]
it can be translated into an action sequence as follows:
\[
[select,layout][x-axis][filter,fill,hide][aggregate,annotate]
\]
with the changes of the corresponding unit visualization shown in Fig.~\ref{fig:teaser}. Semantically, it indicates to (1) \textit{select} a collection of data records, i.e., students, from the data source and represent each record as a unit and layout all the records in the unit visualization; (2) \textit{encode} the attribute, i.e., year of birth,  by x-axis; (3) \textit{filter} out the records that satisfies the condition and highlight them by filling each qualified unit color and (4) \textit{hide} those unqualified ones; (5) \textit{aggregate} the records, i.e., students born in 2000, by counting the units and annotate on records via a tooltip to illustrate the total number of students.

\vspace{-2mm}
\section{Evaluation}
We first estimated the decomposition and calibration network via quantitative experiments and verified the effectiveness of the generated datamation via a controlled user study. Finally, interviews with two expert users were performed to verify the usability of our system. 
\vspace{-2mm}
\subsection{Quantitative Evaluation}
We evaluated the performance of the decomposition network based on the corpus introduced in Section~\ref{sec:decomposition}. 
In particular, we computed the rate of exactly match between the decomposition results and ground truth, and calculated the averaged SARI scores~\cite{xu2016optimizing} of the decomposition results. 
SARI score is commonly used in text simplification tasks, 
it evaluates the goodness of words that are added, deleted, and kept in the simplified sentences.
Two baseline models introduced in~\cite{wolfson2020break} were used for comparison. In particular, the first baseline ($\mathcal{B}_1$) was a sequence-to-sequence neural network with a 5-layer LSTM encoder and a cross attention. The second one ($\mathcal{B}_2$) was another sequence-to-sequence model that incorporated a copy mechanism~\cite{gu2016incorporating} to deal with unseen queries. We trained these models based on the entire corpus and tested their performance via a validation set containing 30\% of data that were randomly selected from the corpus with the data queries being replaced by similar but different questions. The evaluation results are summarized in Table~\ref{tab:decompositionevaluation}.


\begin{table}[htb]
\caption{Performance evaluation of the query decomposition models.}
\setlength\aboverulesep{0pt}
\setlength\belowrulesep{0pt}
\renewcommand{\arraystretch}{1.3}
\centering
\def\arraystretch{1.3}
\begin{tabular}{ccc}
\toprule
\rotatebox{0}{Models} & \rotatebox{0}{Exact Match} & \rotatebox{0}{SARI}
\\\midrule\midrule
    $\mathcal{B}_1$ & 25.64\% & 0.759\\
$\mathcal{B}_2$ & 38.39\% & 0.812 \\
$\mathcal{D}$ & \textbf{82.23\%}  & \textbf{0.876}
\\\bottomrule
\end{tabular}
\label{tab:decompositionevaluation}
\end{table}

\vspace{-3mm}

We measured the effectiveness of the calibration network via \textit{success rate} and the \textit{retain rate}. Here, the \textit{success rate} was defined as the percentage of the incorrect decomposition results $S_d$ that could be successfully amended by the calibration network via the feedback $S_c$ among all the incorrect results. The \textit{retain rate}
was defined as the percentage of correct decomposition results that remain correct after updating the model's parameters via calibrations.  We performed the experiment based on the above decomposition network $\mathcal{D}$ and calculated the \textit{success rate} and the \textit{retain rate} based on the dataset introduced in Section~\ref{section:calibrationnetwork}. In particular, we used 80\% of the data to train the calibration network and used the rest 20\% for validation. As a result the success rate was 76.61\% and the retain rate was 91.79\%. 


\vspace{-2mm}
\subsection{User Study}
We re-performed the experiment (Study-I) described in ~\cite{pu2021datamations} based on our system to verify the effectiveness of the generated datamations. In particular, 40 lab students (18 males and 22 females, mean age 25.1, SD 1.77) from a design college with the background of visual communication design were invited to take part in our study. They were divided into two groups (i.e., $G_1, G_2$), where each group had 20 members. A between-subject study was performed, in which two groups of people were asked to answer the same question by exploring the data via datamations generated by \name and the static charts (the last frames of the datamations), respectively. 
\vspace{-1.5mm}
\paragraph{\textit{Procedure and Tasks}}
During the study, the participants were asked to resolve a paradox as shown in Fig.~\ref{fig:userstudy}: it seems people with higher education (PhD) have a lower average income (Fig.~\ref{fig:userstudy}(c)), but when we take the work fields into consideration, we obtain the opposite result, i.e., people with higher education have a higher income in both industry and academia (Fig.~\ref{fig:userstudy}(d)). The truth is that people with lower education tended to work in industry, where a higher salary is usually paid; and these people outnumbered  the people with higher education by a large margin. We visualized the two opposite sides of the paradox in a pair of datamations generated by \name as illustrated in Fig.~\ref{fig:userstudy}(a,b). We showed these datamations to the participants in $G_1$ and showed the last frames as static charts to the participants in $G_2$. These participants were asked to read these visual representations carefully and then select the truth of the paradox from 8 potential choices (7 were distractors) provided by us (refer to~\cite{pu2021datamations} for details about the choices). Our hypothesis was $G_1$ tends to have higher accuracy than that of $G_2$.


\vspace{-1.5mm}
\paragraph{\textit{Results}} 
As a result, only 40\% of participants in $G_2$ correctly found out the truth of the paradox, while 75\% of participants in $G_1$ did so. Chi-squared test showed that $G_1$ performed significantly better ($\chi^2(1,N=40)$=5.01, p\textless.05, one-tailed) than $G_2$ in the experiment. The gap between $G_1$ and $G_2$ was 35\%, which was a sizeable difference with Cohen’s h-value equals to 0.68. 
This result supported our hypothesis.








\vspace{-0.6em}
\subsection{Interview with Experts}
\begin{figure}[tb]
\setlength{\abovecaptionskip}{10pt}
\centering 
\includegraphics[width=\columnwidth]{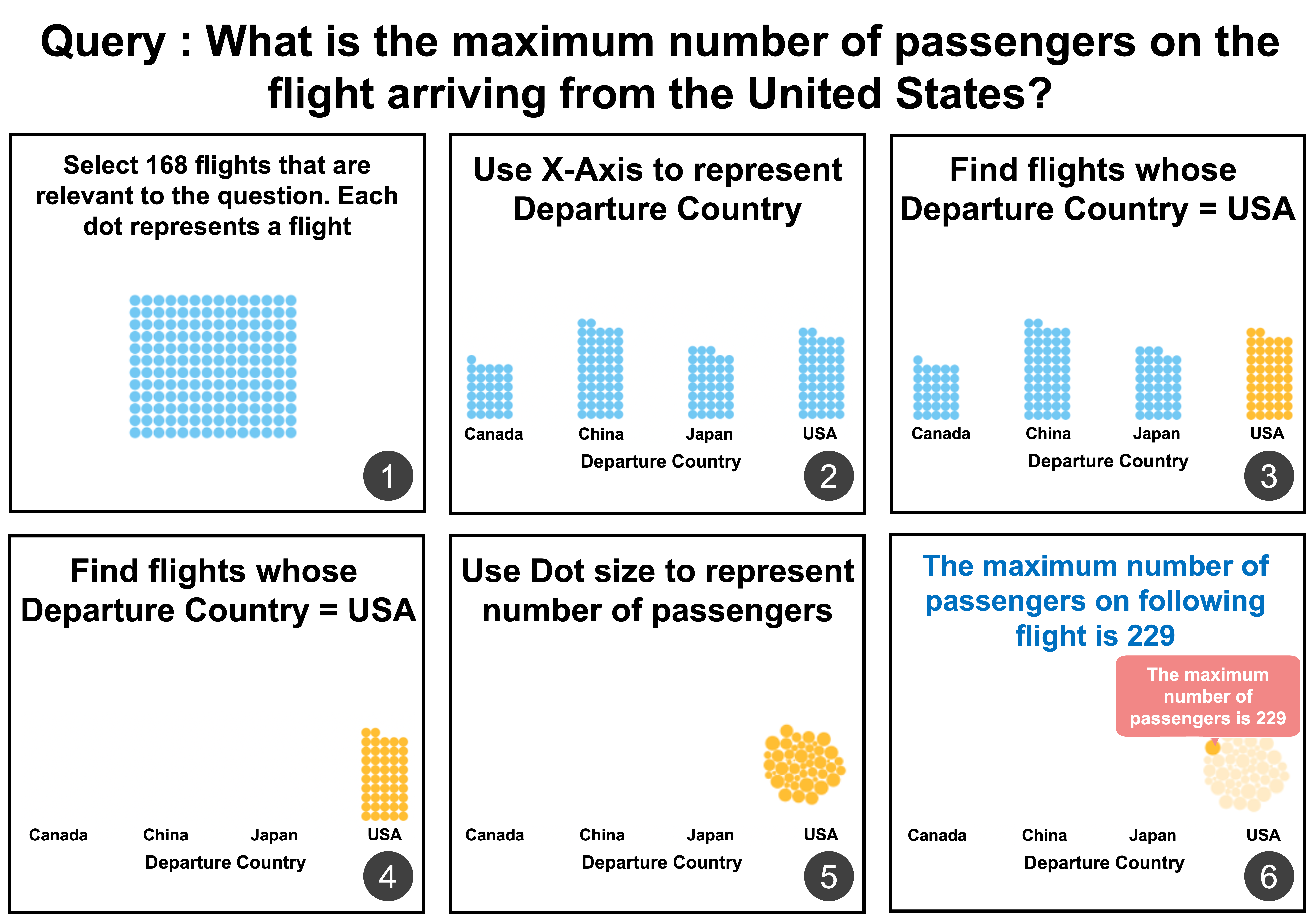}
\vspace{-1.5em}
\caption{The datamation produced by our expert user during the interview that illustrates the analysis pipeline of a dataset about flights landing in Australia in May 2013.}
\label{fig:casestudy1}
\vspace{-1.5em}
\end{figure}
Two expert users were interviewed separately. The first expert ($E1$) was a researcher in data visualization who had published over 10 TVCG papers. The second expert worked for an IT consulting company, whose job was to analyze customer data using tools like PowerBI and Tableau. During the interview, the experts were asked to create datamations using \name based on datasets selected from the spider corpus. They tried a number of data queries and the corresponding datamations were created. Their operations, generation results, and comments were recorded in detail. Each interview lasted for about 1.5 hours. Here, we first demonstrate two example datamations generated by the experts and then report their comments in general.
\vspace{-1mm}
\paragraph{\textit{Case I: Flights Analysis} (\bf{E1}).} The first example was based on a dataset that consisted of 168 international flights landed in Australia in May 2013. Each flight had four attributes: date, flight number, departure country, and the number of passengers.  Fig.~\ref{fig:casestudy1} showed a valid datamation directly generated by \name after the expert input the queried \textit{``what is the maximum number of passengers on the flight arriving from the United States?"} The first frame illustrated all the related flights with each flight visualized as a unit. These flights were grouped by the countries that they arrived from in the second frame. The next three frames gradually highlighted the flights from the USA and encoded the size of each unit by the number of passengers on the corresponding flight. Finally, in the last frame, a tooltip was added to point out the flight had the maximum number of passengers.

\begin{figure}[tb]
\setlength{\abovecaptionskip}{10pt}
\centering 
\includegraphics[width=\columnwidth]{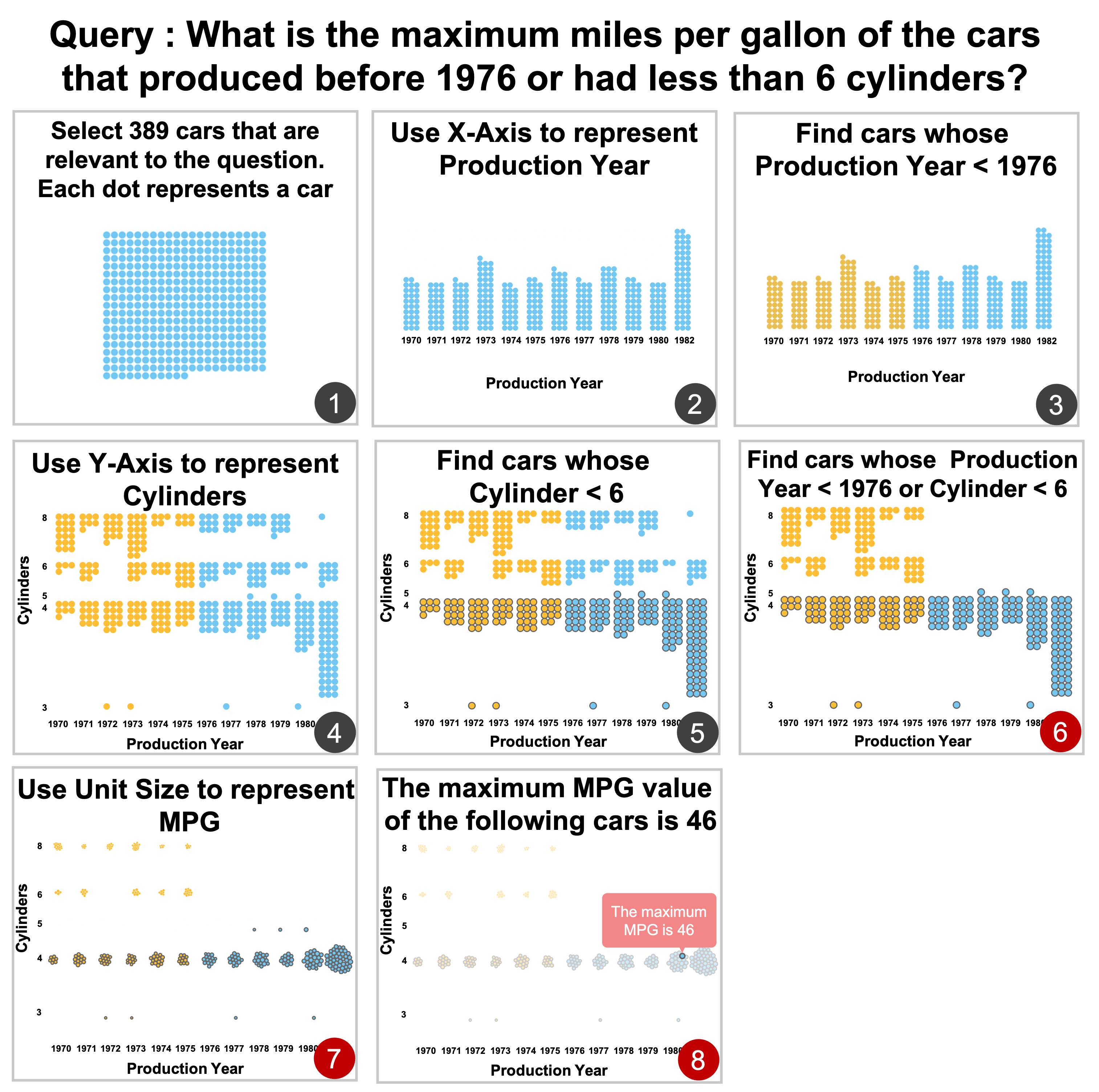}
\vspace{-1.5em}
\caption{The datamation produced by our expert user that illustrates the analysis pipeline of a dataset consisting of 389 cars.}
\label{fig:case2}
\vspace{-2.0em}
\end{figure}
\vspace{-1mm}
\paragraph{\textit{ Case II: Cars Dataset} (\bf{E2}).} The second example as shown in Fig.~\ref{fig:case2} demonstrated a data query with multiple conditions about the famous cars dataset. The query initially resulted in a problematic datamation that contained only the first six frames shown in Fig.~\ref{fig:case2}. The last two frames were added by $E_2$ and used as a feedback to calibrate the model. Specifically, to answer the data query, a set of 389 relevant cars were selected and visualized as units (Fig.~\ref{fig:case2}(1)), which were latter grouped by their production year (Fig.~\ref{fig:case2}(2)) with the targets highlighted in yellow (Fig.~\ref{fig:case2}(3)). After that, these cars were further divided regarding to their number of cylinders (Fig.~\ref{fig:case2}(4)) with the cars with less than 6 cylinders highlighted by a thicker broader (Fig.~\ref{fig:case2}(5, 6)). The last two frames encoded the MPG by size (Fig.~\ref{fig:case2}(7)) and pointed out the final results by a tooltip (Fig.~\ref{fig:case2}(8)).

\vspace{-1mm}
\paragraph{\bf Interview Feedback}
Due to the page limitation, we restrict our report to summarizing only some major feedback from the experts:

\textit{Data query decomposition.} All the experts felt starting a datamation generation process by inputting a nature language query was a ``good idea that greatly reduces the technique barriers.'' $E1$ mentioned: ``resolve a natural language question into data operations is useful for users with little data analysis background.'' $E2$ said: ``although [the system] cannot always provide a comprehensive [decomposition] result, it is always a good start by pointing out the potential directions to solve problem.'' 

\textit{The authoring tool.}
All experts were able to successfully generate datamations based on our system. They were quite excited when seeing an analysis pipeline shown in an animation. ``This is my first time to see such kind of tool,'' said $E2$ and he continued: ``this feature is very nice, I can hardly generate this kind of animations using the tools that I have ever used.'' $E1$ believed that ``showing the analysis via an animation is very intuitive" and ``it is a good strategy for interpreting an analysis result.'' At the same time, both $E1$ and $E2$ praised the interactive authoring functionalities supported in our system and even constructively suggested a number of new features such as ``recommend proper operations for users to choose when creating an analysis pipeline ($E2$)'' and ``provide more visualization styles ($E1$).'' In addition, $E1$ believed that our feedback mechanism is useful as ``it could help the underlying model to reduce errors in the next time.'' $E2$ also mentioned that ``[with the help of feedback] I can make the result better and better.''

\vspace{-2mm}
\section{Conclusion, Limitations, and Future Work}
In this paper, we presented the authoring tool \name, developed for creating datamations. To the best of our knowledge, it is the first tool that supports datamation design and generation. Given a dataset and a question, \name can automatically decompose the question into a sequence of data analysis operators and generate a datamation based on unit visualization. \name also allows the user to modify and edit the generated results. Our user studies showed that \name is highly rated for generating datamations to explain data analysis processes. Its editing function also showed to be effective in correcting the automatically generated results. 

There are a number of limitations of our work that are worth further study in the future. First, both the decomposition and calibration networks are pre-trained models, with therefore limited capability of dealing with unseen data. Second, the current calibration network fixes the decomposition errors, but it is restricted in terms of improving the overall performance of the decomposition model. A knowledge accumulation mechanism that could incrementally capture the knowledge from users' feedback to gradually change the decomposition network is still desired. However, this is a difficult problem known in deep learning that is worth further exploration. Third, the scalability of the visualization design needs to be improved to support larger datasets.

\section*{ACKNOWLEDGMENTS}
Nan Cao and Qing Chen are the corresponding authors. This work was supported in part by NSFC 6200070909, NSFC 62061136003, NSF Shanghai 23ZR1464700, and NSFC 62072338. 

\bibliographystyle{abbrv-doi}
\bibliography{main}
\end{document}